# External control of the direction of magnetization in ferromagnetic InMnAs/GaSb heterostructures

X. Liu,[a] W. L. Lim,[a] L. V. Titova,[a] T. Wojtowicz,[a,b] M. Kutrowski,[a,b] K. J. Yee,[a]
M. Dobrowolska,[a] J. K. Furdyna,[a] S. J. Potashnik,[c] M. B. Stone,[c] P. Schiffer,[c]
I. Vurgaftman,[d] J. R. Meyer[d]

[a]Department of Physics, University of Notre Dame, Notre Dame, Indiana 46556, USA
[b]Institute of Physics, Polish Academy of Sciences, Al. Lotników 32-46, PL-02668 Warsaw, Poland
[c]Department of Physics and Materials Research Institute, Pennsylvania State University, University Park,
Pennsylvania 16802, USA
[d]Code 5613, Naval Research Laboratory, Washington, DC 20375, USA

## Abstract

In this paper, we demonstrate external control over the magnetization direction in ferromagnetic (FM) $In_{1-x}Mn_xAs$/GaSb heterostructures. FM ordering with $T_C$ as high as 50 K is confirmed by SQUID magnetization, anomalous Hall effect (AHE), and magneto-optical Kerr effect (MOKE) measurements. Even though tensile strain is known to favor an easy axis normal to the layer plane, at low temperatures we observe that the magnetization direction in several samples is intermediate between the normal and in-plane axes. As the temperature increases, however, the easy axis rotates to the direction normal to the plane. We further demonstrate that the easy magnetization axis can be controlled by incident light through a bolometric effect, which induces a pronounced increase in the amplitude of the AHE. A mean-field-theory model for the carrier-mediated ferromagnetism reproduces the tendency for dramatic reorientations of the magnetization axis, but not the specific sensitivity to small temperature variations.

Recent experiments have demonstrated that the magnetic properties of the ferromagnetic (FM) heterostructure $In_{1-x}Mn_xAs/GaSb$ can be controlled by an external perturbation, such as illumination [1] or applied electric field [2]. Since the FM interactions in III-Mn-V alloys are mediated by free holes [3], modulating the hole concentration $p$ provides a natural pathway to manipulation of the magnetism. In the present work, we demonstrate that the FM in $In_{1-x}Mn_xAs/GaSb$ heterostructures can also be controlled by light-induced bolometric effects that vary the *direction* of the magnetization. Such an externally-induced rotation of the magnetization could in principle be used to perform operations in quantum-dot-based Q-bit gates.

We grew a number of $p$-type $In_{1-x}Mn_xAs$ ($x = 0.02$-$0.08$) layers by molecular beam epitaxy at 210°C on GaAs substrates using methods similar to those described in Refs. [4-6]. In the present work, we find that the details of the growth procedures and conditions have a dramatic influence on the magnetic properties (i.e., easy axis of magnetization) of the $In_{1-x}Mn_xAs/(Al,Ga)Sb$ heterostructures. Under optimal conditions, the reflection high-energy electron diffraction (RHEED) showed a (1×1) or (1×2) reconstruction. All of the samples reported here had the structure InMnAs/GaSb(700nm)/AlSb(150nm)/GaAs. The InMnAs film was between 10 and 15nm thick, and its interface with the GaSb layer had an InSb-like bond type. While Hall measurements at 300 K implied free hole densities ($p$) in the 1-2×$10^{19}$cm$^{-3}$ range, those data are probably dominated by carriers in the GaSb layers, since at least an order of magnitude higher $p$ are required to account for our observation of relatively large $T_C$ ($\geq 30$

K). Alternatively, Hall measurements on InMnAs/AlSb structures with similar $T_C$ yielded $p \approx$ 1-3×10$^{20}$cm$^{-3}$, which is much more reasonable.

To examine the magnetic properties of the InMnAs films, magnetization ($M$) measurements were carried out as a function of temperature using a dc superconducting quantum interference device (SQUID) magnetometer, with the magnetic field $\boldsymbol{H}$ applied either normal to the layer plane ($\boldsymbol{H}$‖[001]) or in the plane ($\boldsymbol{H}$‖[110]). For $|\boldsymbol{H}|$ = 10 Gauss, Figs. 1(a) and 1(b) show temperature-dependent magnetization results for two samples. In addition, the insets of Fig. 1 show hysteresis loops at $T$ = 5K for both [110] and [001] field directions. These data clearly indicate that there are *two types* of In$_{1-x}$Mn$_x$As films, whose easy and hard axes are oriented differently at low temperatures. The $\boldsymbol{H}$‖[001] results in Fig. 1(a) indicate that the easy axis for Sample 1 ($x$ = 0.07, $T_C$ = 42K) is perpendicular to the layer plane. On the other hand, it appears from the $\boldsymbol{H}$‖[110] data in Fig. 1(b) that the easy axis for Sample 2 ($x$ = 0.05, $T_C$ = 30K) lies in the plane at low $T$, although it will be shown that the actual orientation is at some intermediate angle. For each sample, a larger magnetization is observed for the easy axis, and the insets to both figures show more pronounced (nearly square) hysteresis loops for the easy axes. The data for Sample 2 clearly contradict the usual expectation that the InMnAs film's easy axis should lie perpendicular to the layer plane, due to the tensile strain induced by a lattice mismatch of $\approx$ 0.78% between the InMnAs magnetic layer and the relaxed GaSb buffer [4]. Anomalous Hall effect (AHE) and magneto-optical Kerr effect (MOKE) measurements were also performed, both of which showed analogous sensitivities to each given sample's easy magnetization axis.

Figure1(b) shows that while the ***H***||[001] magnetization for Sample 2 increases with temperature in the 10-18 K range, the ***H***||[110] magnetization dominates at $T \rightarrow 0$ but then drops toward zero at $T$ far below the Curie temperature. We attribute this behavior, which is opposite to that observed by Sawicki *et al.* for GaMnAs [7], to spin reorientation, *i.e.*, the easy axis flips from non-perpendicular to perpendicular orientation as the temperature increases. In order to directly trace the reorientation of the spontaneous magnetization, we examined the temperature dependence of the remanent magnetization for Sample 2 along selected directions, using the following procedure. After cooling to 10K in zero field, the sample was magnetized in the [001] direction at |***H***| at least a few times the coercive field (*e.g.*, 1000Gauss). The field was then removed at 10 K, and the sample rotated by various angles θ toward the [110] direction (*e.g.*, θ = 0°, 45°, -45°). The remanent magnetization component along these directions was measured as a function of increasing temperature, taking care to remove any background magnetization due to possible MnAs or MnSb inclusions. The results shown in Fig. 2(a) clearly indicate that even though the sample was magnetized in the [001] direction, the magnetization rotates to a direction intermediate (approximately half-way) between the [001] and [110] directions after the field is removed. However, within the relatively narrow window of temperatures between 10 and 18K the magnetization *reorients* to the direction normal to the plane. Further evidence for the temperature dependent spin reorientation in Sample 2 is provided by the hysteresis loops shown in Fig. 2(b) for ***H***||[001]. As $T$ increases from 10 K to 20 K, the loop evolves from an elongated to a nearly square profile, accompanied by an increase of the remanent magnetization.

The data in Figs. 1 and 2 suggest that temperature provides a convenient means for controlling the magnetization orientation in a subset of the InMnAs/GaSb heterostructures. We demonstrated this by using illumination to tune $T$ via a bolometric effect. A beam of white light from a standard 100W bulb was focused onto the InMnAs sample mounted on the cold finger of a closed-cycle helium optical cryostat. The sample temperature was monitored by two thermo-resistors attached to the cold finger above and below the sample, and $T$-dependent Hall measurements were carried out with and without illumination. For Sample 3 ($x = 0.05$, $T_C$ = 35K), whose magnetic properties are similar to those of Sample 2 (see inset in the figure), the solid curve in Fig. 3 shows the magnetization deduced from AHE (assuming M $\propto$ $\rho_{xy}/\rho_{xx}$, where $\rho_{xy}$ and $\rho_{xx}$ are Hall resistivity and sheet resistivity, respectively) at 10K in the dark. The dashed curve indicates that illuminating the light produces a pronounced *increase* in the amplitude of the AHE (by as much as a factor of three), accompanied by the change to a nearly-square hysteresis. The reverse effect is observed when the light is turned off.

Sawicki *et al.* [7] suggested that the spin reorientation observed in their experiments reflected the spin anisotropy of the valence subbands in a zinc-blende semiconductor [8,9], which can depend on the hole and Mn concentrations, epitaxial strain, and temperature. It should be emphasized, however, that the In$_{1-x}$Mn$_x$As/GaSb case is even more complicated. To investigate this quantitatively, we carried out mean-field-theory simulations of the carrier-mediated ferromagnetism in conjunction with the 8-band effective bond-orbital method (EBOM), including strain effects within deformation-potential theory [10]. We first consider the typical parameters: 1.0 eV for the exchange integral, $p$ = 1.5$\times$10$^{20}$cm$^{-3}$, and $x$ = 0.05. In this and subsequent calculations, the Mn concentration was adjusted at each hole density so as to reproduce the observed $T_C$ of $\approx 30$ K. Allowing for arbitrary magnetization direction, the calculated spin anisotropy implies an easy axis that lies at an angle intermediate between the in-plane [110] direction and the out-of-plane [001] direction (41° from [110], toward [001]), irrespective of temperature. On the other hand, even a relatively small variation of the hole density can produce a dramatic reorientation of the easy axis. For example, increasing the hole density to $\approx 2 \times 10^{20} cm^{-3}$ causes the low-$T$ easy axis to shift to the [001] direction. Surprisingly, increasing the hole density further, to $3.5 \times 10^{20} cm^{-3}$, induces a switch in the easy axis back to the layer plane, but now along [100] rather than [110]. Thus the simulations confirm the potential for dramatic reorientations of the magnetization direction, although it remains unclear why the easy axis should be so sensitive to small temperature variations (within a rather narrow window of $\Delta T \approx 8$ K).

One interpretation is that the data reflect a variation of the hole density with $T$. Whereas such a variation is unlikely in an isolated InMnAs film, the broken-gap type-II band alignment of the InMnAs/GaSb heterojunction presents a more complicated situation. Some of the holes originating in the InMnAs are transferred to the adjacent GaSb layer due to its higher valence band maximum, thereby creating both an internal electric field and a density gradient in the InMnAs layer. Further study is necessary to clarify the temperature-dependent process, however, since the screening length in InMnAs is very short ($\approx 10$ Å) and it appears that rather fortuitous parameters are required in two different samples. Similar considerations were invoked by Munekata *et al.* to explain the enhancement and control of magnetic processes by light illumination in the same InMnAs/GaSb system [3,11]. Other possibilities include a variation of the net strain with temperature, or a sensitivity of the carrier-mediated ferromagnetism to either the hole scattering properties or the spatial distributions of the hole wavefunctions in the presence of very high defect densities.

This work was supported by the DARPA/SpinS Program through the Office of Naval Research.

**Figure Captions**

Fig. 1 SQUID data for the $In_{1-x}Mn_xAs/GaSb$ magnetization $M$ as a function of temperature $T$ at $|\boldsymbol{H}|$ = 10G (main plot) and *vs.* applied magnetic field $H$ (insets, not corrected for demagnetization) for: (a) Sample 1 and (b) Sample 2. The magnetic field is applied in either the [110] (open circles) or [001] (filled points) direction. Note that the easy axis is different for the two samples. Magnetoresistance (insert open box symbol here) data are also shown at $|\boldsymbol{H}|$ = 0G for the easy axis of each sample.

Fig. 2 (a) Remanent magnetization in Sample 2 as a function of temperature for three crystalline axes. While the sample was magnetized in the [001] direction at $T$ = 10K, the easy axis of magnetization clearly switches from a direction between [001] and [110] to the [001] direction on warming up. (b) Magnetization as function of applied magnetic field for Sample 2, measured at $T$ = 10K and 20K by SQUID.

Fig. 3. Magnetization curves deduced from AHE, for Sample 3 at $T$ = 10 K with and without illumination. The inset plots the same sample's magnetization at $|\boldsymbol{H}|$ = 10Gauss as a function of temperature. In all cases, the magnetic field is applied normal to the layer plane. The dashed curve is similar to the magnetization curve measured between 15K and 20K without light illumination (not shown here).

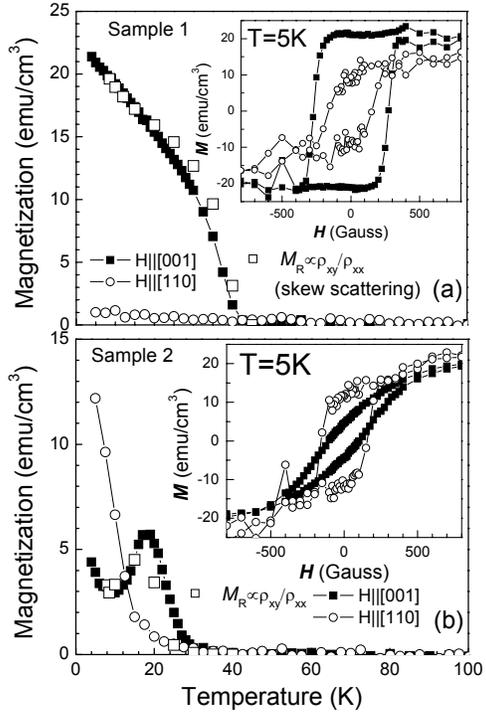

Fig. 1 X. Liu *et al.*

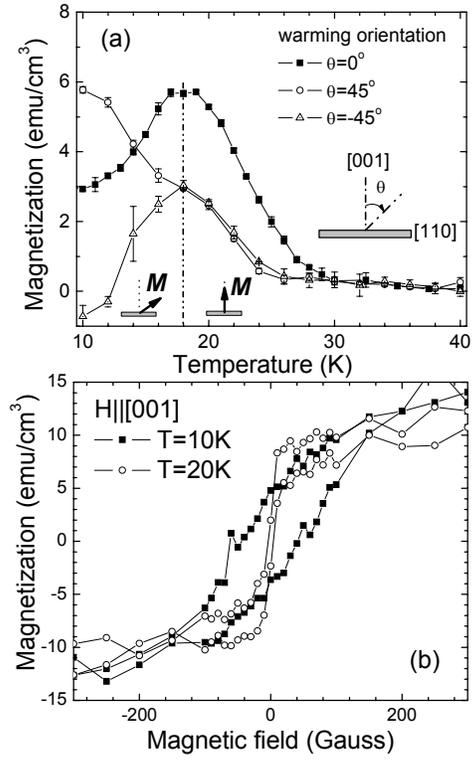

Fig. 2 X. Liu *et al.*

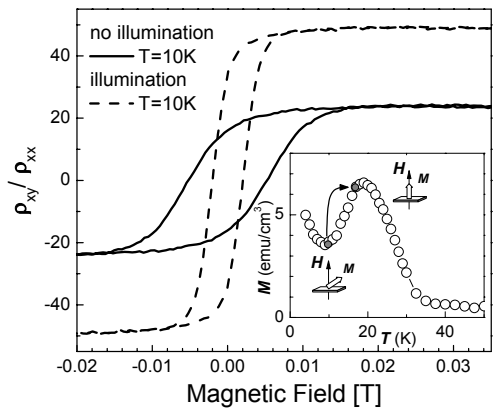

Fig. 3 X. Liu *et al.*